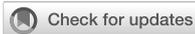

# Mapping acceptance: micro scenarios as a dual-perspective approach for assessing public opinion and individual differences in technology perception

Philipp Brauner*

Communication Science, RWTH Aachen University, Aachen, Germany

Understanding public perception of technology is crucial to aligning research, development, and governance of technology. This article introduces micro scenarios as an integrative method to evaluate mental models and social acceptance across numerous technologies and concepts using a few single-item scales within a single comprehensive survey. This approach contrasts with traditional methods that focus on detailed assessments of as few as one scenario. The data can be interpreted in two ways: Perspective (1): Average evaluations of each participant can be seen as individual differences, providing reflexive measurements across technologies or topics. This helps in understanding how perceptions of technology relate to other personality factors. Perspective (2): Average evaluations of each technology or topic can be interpreted as technology attributions. This makes it possible to position technologies on visuo-spatial maps to simplify identification of critical issues, conduct comparative rankings based on selected criteria, and to analyze the interplay between different attributions. This dual approach enables the modeling of acceptance-relevant factors that shape public opinion. It offers a framework for researchers, technology developers, and policymakers to identify pivotal factors for acceptance at both the individual and technology levels. I illustrate this methodology with examples from my research, provide practical guidelines, and include R code to enable others to conduct similar studies. This paper aims to bridge the gap between technological advancement and societal perception, offering a tool for more informed decision-making in technology development and policy-making.

# 1 Introduction

Technological advancements are often accompanied by dilemmas and they must be aligned with human norms and values. History has many instances of such ethical dilemmas, such as mechanization and industrialization, leading to enhanced productivity but also accompanied by substandard working conditions (Engels, 1845; Watt, 1769), movable types and the printing press yielding increased literacy but resulting in the dissemination of pamphlets containing misinformation (Steinberg, 1974; Eisenstein, 1980), and the invention of clothing for protection and warmth leading to the environmental repercussions of fast fashion, causing ecological damage (Kvavadze et al., 2009; Niinimäki et al., 2020).

When technologies become a part of our life, it is essential to integrate the perspective of us—the people—to understand how we evaluate them, what we attribute to them, and how they relate to our norms and values (Guston and Sarewitz, 2002; Rogers et al., 2019; Lucke, 1995). When technologies reflect peoples' values, they are more likely to be accepted, adopted, and integrated into daily life. Conversely, if a technology conflicts with prevailing values, it may face resistance or rejection. However, technology may change our norms and values and our norms and values may shape how a technology is used. For instance, the Internet has fostered values of openness and connectivity, while these values have, in turn, driven the development of social media platforms. Similarly, technologies can afford new possibilities that lead to the development of new values. For example, the rise of renewable energy technologies has spurred values around environmental sustainability. However, there are instances where technologies and values are in opposition. Surveillance technologies, for example, clash with values of privacy and individual freedom. Also, technologies often introduce ethical dilemmas where existing values are challenged, such as the advent of genetic editing technologies like clustered regularly interspaced short palindromic repeats (CRISPR) raises questions about the value of human life and natural processes.

There are various methods for assessing peoples' perception of technologies: ranging from scenario-based approaches, over living labs, to hands-on experiences with readily available technologies (Tran and Daim, 2008; Grunwald, 2009). The majority of empirical approaches use different concepts of technology acceptance to assess specific technologies and systems. Referring to model-based approaches, the constructs behavioral intention to use and actual use are often applied to measure technology acceptance (Davis, 1989; Marangunić and Granić, 2015). Other approaches focus more on affective evaluations, addressing the social perception of specific technologies and systems (Agogo and Hess, 2018; Zhang et al., 2006). Furthermore, the evaluation of single technologies often contains a modeling and trade-off between specific technology-related perceived (dis-)advantages affecting the final evaluation and acceptance (Buse et al., 2011; Offermann-van Heek and Ziefle, 2019).

Although research on technology acceptance and evaluation has increased significantly in the last decades, the majority of the studies focus on the evaluation of single applications (Rahimi et al., 2018; Al-Emran et al., 2018) describing specific requirements, benefits, and barriers of its usage in depth. In contrast, a broader view on diverse technologies' assessment enabling a comparison and meta-perspective on a variety of technologies has rarely been realized so far. Further, most evaluations based on conventional acceptance models or their adaptations do not facilitate mapping or contextual visualization of a wider range of technologies and concepts.

Therefore, this article aims at presenting a novel micro-scenario approach, enabling a quantitative comparison of a broad variety of technologies, applications, or concepts based on affective evaluations, in parallel with the interpretation of an individual's assessment as individual dispositions, as well as a concept of visualizing the evaluations as visual cognitive maps.

The article is structured as follows: Section 1 provides the introduction and motivates this methodological approach. Section 2 reviews the current state of technology acceptance evaluations and related measures, highlighting existing research gaps. Section 3 defines micro-scenarios as an integrated contextual perspective and discusses the strengths and limitations of this approach. Section 4 introduces guidelines and requirements for designing surveys based on micro-scenarios. Section 5 presents a concrete application example, showcasing the results of a recent study on the acceptance of medical technology. This example demonstrates the practical value of the approach and the insights it can provide (all data and analysis code are available as open data). Section 6 concludes with a summary and a discussion of the methodological strengths and limitations of the approach, as well as its overall usefulness. Finally, the Appendix details the technical implementation of micro-scenario-based surveys, along with actionable examples and R code for conducting similar studies.

## 2 Background and related measures

The following section presents the theoretical background and introduces related empirical concepts and approaches, as well as related methodological procedures.

## 2.1 Related concepts and approaches

A fundamental concept in acceptance research is mental models. These are simplified, cognitive representations of real-world objects, processes, or structures that enable humans and other animals to evaluate the consequences of their (planned) actions. These simplified models influence our behavior (Jones et al., 2011; Johnson-Laird, 2010; Craik, 1943): When aligned with reality, they facilitate efficient and effective interactions with the surroundings (Gigerenzer and Brighton, 2009). Conversely, erroneous mental models restrict the correct assessment of the environment and hinder accurate inferences (Gilovich et al., 2002; Breakwell, 2001).

Extracting mental models through empirical research provides insights into how basic attitudes and attributions are shaped and change.

For this purpose, many qualitative (for example, interviews and focus groups or rich picture analysis) and quantitative approaches (for example, surveys or experimental studies) are available. One frequently used method in acceptance research involves scenarios depicting technologies or their applications, which are integrated in qualitative, quantitative, or mixed-methods approaches (Kosow and Gaßner, 2008). In this approach, a new technology or service is described textually and/or visually within a scenario and then evaluated by study participants based on various criteria. Typically, these scenarios are designed to let participants evaluate a single technology, application, or situation in detail. Only occasionally, a few (rarely more than three) different technologies or their applications are assessed. Through these scenarios, participants evaluate their perceptions, attitudes, and acceptance of the specific research object. While these responses are not the mental models, they reflect the participants mental models.

There are multiple ways to describe the perception of technologies and the influencing factors involved. A prominent example are studies based on the technology acceptance model (TAM) or the increasingly specific models derived from it (Davis, 1989; Rahimi et al., 2018). TAM postulates that the later actual use of a technology—originally office applications—can be predicted in advance via a model of the individuals' perceived ease of using the system, and the perceived usefulness, and the intention to use the system. Later models have extended the concept of predicting the later use through the usage intention and an increasingly diverse set of antecedents. Examples include the hedonic value of a product, or if others could provide support in case of troubles (Venkatesh et al., 2012). Nowadays, new models are being proposed for each seemingly new technology; but rarely are different technologies compared in a single study. While the core idea remains the same—predicting use by linking intention to use to other factors—there are now many an overwhelming number of models and constructs used in technology acceptance research (Marikyan et al., 2023 gives a meta-review on the constructs used in 693 studies).

As not every technology is *used* by individuals (such as a nuclear power plant), other models focus on other outcome variables. For example, the value-based acceptance model shares many similarities with the TAM (Kim et al., 2007), yet it focusses on a *perceived value* of the evaluated entity instead of the *intention to use* (and *use*). Again, different predictors are related to the valence as the target variable and researchers can weight the factors that influence to higher or lower valence of a topic.

A common feature of all these approaches is that one or very few technologies or scenarios are assessed in detail. In contrast, the micro-scenario approach looks at many different scenarios and tries to put them in relation to each other and to uncover connections and differences between the scenarios.

Beyond the need to better understand technology attributions and acceptance at both technological and individual levels, there is also a need to enhance our methodological tools. Studies suggest that questionnaires assessing technology acceptance (and likely other questionnaires) may be biased due to the lexical similarity of items and constructs (Gefen and Larsen, 2017). A significant portion of the TAM can be explained solely through linguistic analysis and word co-occurrences (although subjective evaluations further improve the model). To further develop and validate our methods, it is essential to consider different and new perspectives on the phenomena we study (Revelle and Garner, 2024).

## 2.2 Related methods

This section presents existing and partly related methodological procedures in order to identify similarities, but also differences and gaps, the approach presented here addresses.

### 2.2.1 Vignette studies

At first sight, vignette studies are related to this approach, although they are rather the opposite of the method presented here. Vignette studies are a way to find out which characteristics influence the evaluation of people, things, or services. Essentially, in vignette studies, a base scenario is parameterized using certain dimensions of interest, displayed and evaluated by subjects based on one or more evaluation dimensions. Examples include studies on the influence of cognitive biases in evaluating job applications: The same job applications may be framed by the applicants' age, ethnicity, or social group and as target variable, for example, the likelihood of interviewing the person for a job is measured (Bertogg et al., 2020). This approach enables to examine which factors have an influence on, for example, the likelihood to get invited to the job interview and also to quantify the weight of each factor using, for example, linear regressions on the factor that constitute the vignettes (Kübler et al., 2018). The key difference between the established vignette studies and the approach presented here is that vignette studies aim at identifying influencing factors for one particular entity (e.g., an applicant) while micro-scenarios address the influencing factors of different topics in one shared research space.

### 2.2.2 Conjoint analysis

There is also a similarity to the conjoint analysis (CA) approach. CA were developed in the 1960s by Luce and Tukey (1964) and are most prevalent in marketing research. Participants are presented a set of different products that are composed of several attributes with different levels. Depending on the exact methods, they either select the preferred product out of multiple product configurations, or decide whether they have a purchase intention for one presented option. CA results in a weighting of the relevant attributes for production composition (e.g., that car brand may be more important than performance or color) and the prioritizations of the levels of each attribute (e.g., that red cars are preferred over blue ones). While this approach shares some similarities with the micro-scenarios (e.g., systematic configuration of the products resp. scenarios) there are also differences. A key difference is that CA has one target variable (e.g., selection of the preferred product), whereas the micro-scenarios have multiple target variables and each scenario is evaluated. Furthermore, CA has tools for calculating optimal product configurations and market simulators. While the market simulation allows a comparison of multiple actual or fictitious products, it does not facilitate the identification of blank areas in a product lineup or how the products relate to each other beyond a unidimensional preference. Also, while results from a CA can be used to define customer segments by means of a latent class analysis, the individual preferences can not easily be interpreted as personality factors.

### 2.2.3 (Product) positioning

Another similar approach is "positioning" in marketing (Ries and Trout, 2001), in which products and brands in a segment are evaluated in terms of various dimensions and presented graphically. Based on the graph, new products or brands can be developed to fill gaps or reframed and thus moved to different positions. However, the approach presented here does evaluate and map topics. It focuses on an understanding of the public perception of topics, it does not aim to create new topics, and the evaluated topics can usually not easily be changed (i.e., power plant technologies). Furthermore, beyond the positioning, it does not aim at modeling or explaining the role of individual differences in the evaluations.

### 2.2.4 Psychometric paradigm of risk perception

There are similarities between the micro-scenario approach and Slovic's psychometric paradigm and his seminal works on risk perception (Slovic, 1987; Fischhoff, 2015). Based on the analysis of individual studies, his work suggests that risk attributions have a two-factor structure, with dread risk and unknown risk identified by factor analysis. He used these two-dimensional factors to map a variety of different hazards on a scatterplot ("cognitive map") that looks very similar to the visual outcomes of the micro-scenario approach. However, Slovic's approach focusses more on the psychological aspects of how people perceive and categorize risks and it's based on many individual studies. In contrast, micro-scenarios are more pragmatic and allow arbitrary evaluation dimensions. Building on a single integrated survey and considering risk, utility, or other relevant dimensions can inform researchers, decision-makers, and policy makers in a tangible and applicable manner.

### 2.2.5 Experimental factorial designs

A common theme in psychological research is factorial designs that involves manipulating two or more independent variables simultaneously to study their combined effects on one or more dependent variables (Montgomery, 2019; Field, 2009). It allows us to examine and weight the influence of the factors and the interaction effects between multiple factors (Montgomery, 2019). This concept is extensively and predominantly used in experimental cognitive and behavioral research. However, its application in scenario-based acceptance studies is limited. When used in such studies, they typically only involve single factors due to the large number of dependent variables queried, which would otherwise make the surveys unmanageable.

## 2.3 Similarities and methodological gap

Summarizing these different methodological approaches, they indeed share similarities with the approach presented here. However, they differ in terms of the usage context and purpose of use, variable reference and scope, their target size and their comparability. What is still needed is a broader view of the assessment of on diverse technologies, enabling a comparison and meta-perspective on a variety of technologies enabling comparative mappings or visualizations.

Therefore, a novel micro-scenario approach is introduced in the following section. In the single survey, this approach allows both the assessment and comparison of a wide range of topics, applications, or technologies, as well as the measurement of individual differences in the assessments based on affective evaluations.

# 3 Micro-scenarios as an integrated contextual perspective

The goal of the micro-scenario approach is to gather the evaluation of a *wide range of topics or technologies* on *few selected response variables* and put the different evaluations into context. Hereto, the subjects are presented a large number of different short scenarios and how they evaluate those scenarios is measured using a small set of response variables. The scenario presentation can be a short descriptive text, and/or images, or, in extreme cases, just a single word about an evaluated technology or concept. The former offers the possibility to give some explanation on each of the evaluated topics, whereas the latter essentially measures the participants' affective associations toward a single term. Section 4.1 outlines guidelines for creating the set of scenarios.

Each scenario is then evaluated on the same small set of response items. Which dimensions are used for the assessment depends on the specific research question and may, for example, be risk, benefit, and overall evaluation of a technology to identify (in-)balances in risk-utility tradeoffs (cf. Fischhoff, 2015), the intention to use and actual use of technology as in the TAM (cf. Davis, 1989) to identify different motives for using software applications, the perceived sensitivity of data types and the willingness to disclose the data to others to understand the acceptance barriers to personal life-logging and monitoring at the work-place (cf. Tolsdorf et al., 2022), or other dependent variables that match the research focus. I suggest the use of only single item-scales and only to measure the most relevant target dimensions (Fuchs and Diamantopoulos, 2009; Ang and Eisend, 2017; Rammstedt and Beierlein, 2014). Typically, one would use three to five items for the evaluation of each micro-scenario. On the one hand, this sacrifices the benefits of psychometric scales with high internal reliability. On the other hand, this offers the benefits that (a) each scenario can be evaluated quickly and cost-effective (Woods and Hampson, 2005; Rammstedt and Beierlein, 2014), (b) perceived repetitiveness of psychometric scales is avoided and the survey can be more interesting for the participants, and (c) many scenarios can be evaluated in a single survey. Section 4.2 details the selection of suitable items. Figure 1 illustrates this concept.

With a suitable combination of scenarios and dependent variables, the approach offers two complementary research perspectives:

*Perspective 1:* As the first research perspective, the evaluations can be understood as user variables (individual differences between the participants) and correlations between age, gender, or other user factors can be investigated. The evaluation of various topics can essentially be considered as a repeated reflexive measurement of the same underlying latent construct (see Figure 2).

*Perspective 2:* As the second research perspective, the evaluations serve as technology evaluations and relationships between the evaluation dimensions across the different topics can be studied (differences and communalities between the queried topics) (see Figure 3).

This approach has three distinct advantages:

*Efficient evaluations:* One advantage lies in a pragmatic and efficient evaluation of the topics by the participants, as the cognitive effort required to evaluate the topics is comparably low. Following the mainstream model or answering survey items participants (1) need to understand the question, (2) gather relevant information from long-term memory, (3) incorporate that information into an assessment, and (4) report the resulting judgment (Tourangeau et al., 2000). Here, the respondents have to retrieve their attitude toward each topic only once and then evaluate it on a repeating set of the same response items that should be presented in the same order. While the number of items in these studies is high, its repetitive structure responds to them cognitively easily. That facilitates assessing large number of topics within the same survey.

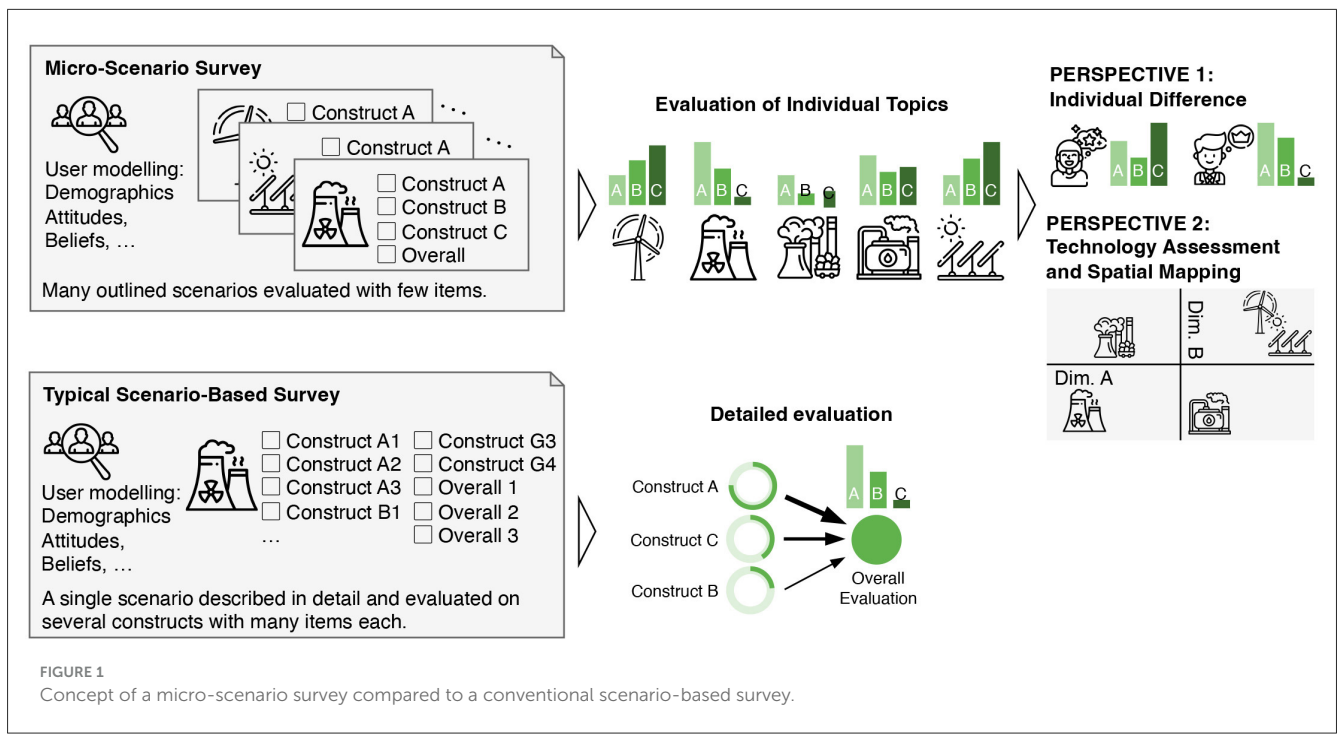

FIGURE 1
Concept of a micro-scenario survey compared to a conventional scenario-based survey.

*Joint evaluations:* In addition, a large number of different topics can be analyzed in a single integrated study. Based on the selected dependent variables for the topics, the relationships among these can further be studied. In a study, we used a linear regression analysis to study the influence of perceived risk and perceived utility on the overall valence of medical technology (Brauner and Offermann, 2024, see the example in Section 5). Based on the calculated regression coefficients and with a high explained variance ($\gg$90%), we could argue that the variance in overall evaluation of medical technology is mostly determined by the perceived benefits rather than the perceived risks.

*Visual interpretation:* Furthermore, the multivariate scenario evaluations can be put into context and presented on two-dimensional spatial maps enabling a visual interpretation of the findings (see Figure 4 for an abstract example and Figure 3 for a view on the required data structure). This representation facilitates the analysis of the spatial relationships between the topics and the identification of topics that diverge from others and thus require particular attention by the public, researchers, or policymakers. To stay in the aforementioned example, we mapped the risk-utility tradeoff across a variety of different topics (see Figure 6 in Section 5). This *visual mapping* of the outcomes can then be interpreted as follows: *First*, one can interpret the breadths and position of the distribution of the topics on the *x*- or *y*-axis. A broader distribution suggests a more diverse evaluation of the topics, whereas a narrow distribution is an indicator for a rather homogenous evaluation. The mean of the distribution of the topics indicates if the topics are—on average—perceived as useful or useless. *Second*, the slope and the intercept of the resulting regression line can be interpreted: The steepness of the slope indicates the tradeoffs between the two mapped variables. If the slope is +1, an increase by one unit of perceived utility means an increase by one unit of perceived risk. Steeper or flatter slopes indicate different tradeoffs. *Third*, one can inspect the position of the individual topics on the map. Elements from left to right are perceived as having less or more risk. Elements from the top to bottom are perceived as having higher to lower utility. Consequently, elements on or near the diagonal are topics where risk and utility are in balance. While some topics are perceived as more and others as less risky, the additional perceived risk is compensated by additional utility. However, if elements are far off the diagonal, there is perceived risk and the utility is unbalanced, potentially because a minor utility does not compensate for a higher degree of risk. Hence, these topics require particular attention from individuals, researchers, or policymakers.

Obviously, other research questions may build on different pairs of dependent variables to be mapped, such as intention and behavior, the same dependent variable by different groups, such as experts and laypeople, or usage contexts, such as passive and active use of technology.

In summary, the micro-scenario approach captures the individual participants' attributions toward various topics but instead of considering these only as individual differences, they are also interpreted as technology attributions and analyzed accordingly.

Consequently, I define micro-scenarios as a methodological approach that facilitates the comprehensive assessment of numerous technologies or concepts on few response items within a single survey instrument. This method enables the quantitative analysis and visual illustration of the interrelationships among the technologies or concepts being investigated. Furthermore, micro-scenarios enable the interpretation of the respondents' overall attributions as personal dispositions, thereby providing insight into individual perceptions and beliefs.

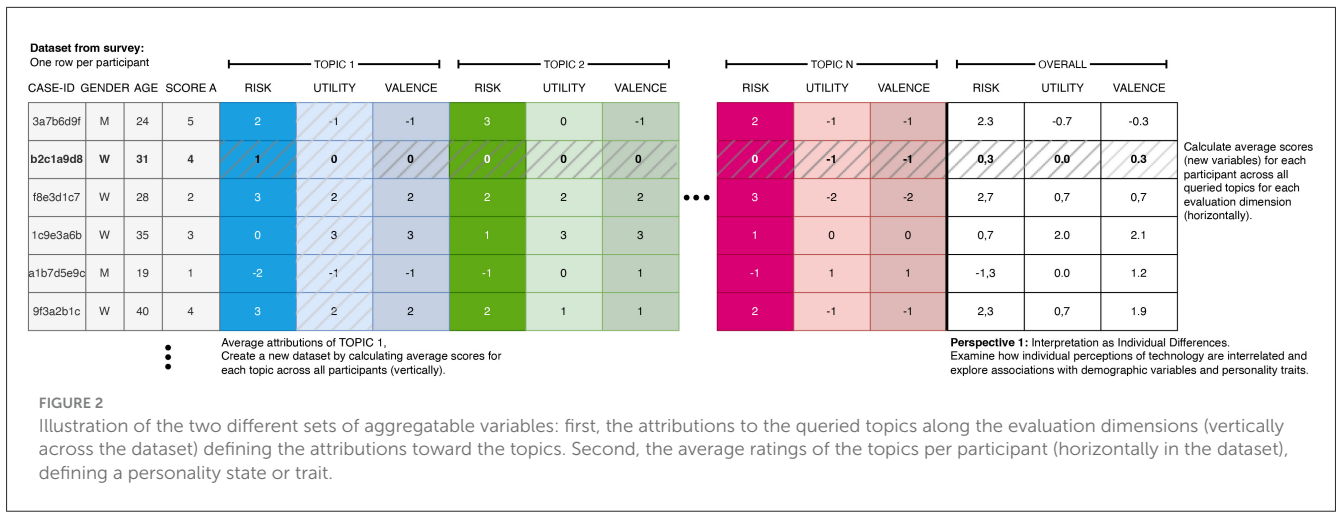

FIGURE 2
Illustration of the two different sets of aggregatable variables: first, the attributions to the queried topics along the evaluation dimensions (vertically across the dataset) defining the attributions toward the topics. Second, the average ratings of the topics per participant (horizontally in the dataset), defining a personality state or trait.

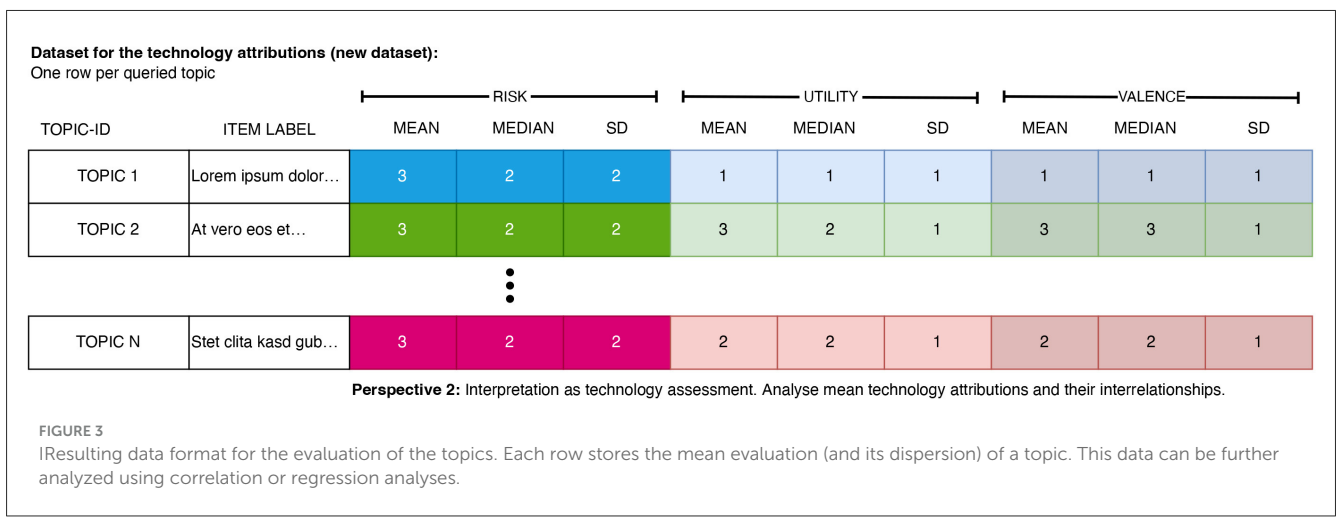

FIGURE 3
IResulting data format for the evaluation of the topics. Each row stores the mean evaluation (and its dispersion) of a topic. This data can be further analyzed using correlation or regression analyses.

# 4 How to conduct micro-scenario surveys

This section outlines the guidelines for conceptualizing a micro-scenario study. Hereby, three areas have to be considered. First, the identification and definition of a suitable research space. Second, the definition of suitable dependent variables that are relevant, suitable for visual mapping, and facilitate further analyses. Finally, the identification of additional variables for modeling the participants that can then be related with the aggregated topic evaluations. In the following, I discuss each point briefly and provide a few suggestions. Obviously, this can neither replace a text book on empirical research methods (e.g., Döring, 2023; Groves et al., 2009; Häder, 2022) nor a systematic literature review on current research topics. However, it should give some guidance on which aspects need to be considered to create an effective survey.

## 4.1 Defining the scenario space

Researchers first have to define the general research domain (such as the perception of Artificial Intelligence, medical technology, or energy sources). Technologies that serve similar functions or are used in similar contexts can be compared in terms of public perception and value alignment. For example, different renewable energy technologies can be compared based on values related to environmental impact and sustainability. However, technologies serving fundamentally different purposes may be less comparable and thus the micro-scenario approach is then not suitable: Comparing an entertainment technology like virtual reality to a healthcare technology like MRI machines may not yield meaningful insights due to the divergent values and expectations involved. Based on this, a set of suitable scenarios needs to be identified. To compile the set of scenarios there are two different approaches:

On the one hand, the scenarios can be defined intuitively, based on the results of an extensive literature review, or as a result of appropriate preliminary studies [such as interviews or focus-groups (Courage and Baxter, 2005)]. However, this bears the risk that the selection of queried topics is neither random nor systematically constructed. While the analysis can yield interesting results, there is a risk that the findings may be affected by a systematic bias (for example, Berkson, 1946's paradox, where a biased sample leads to spurious correlations).

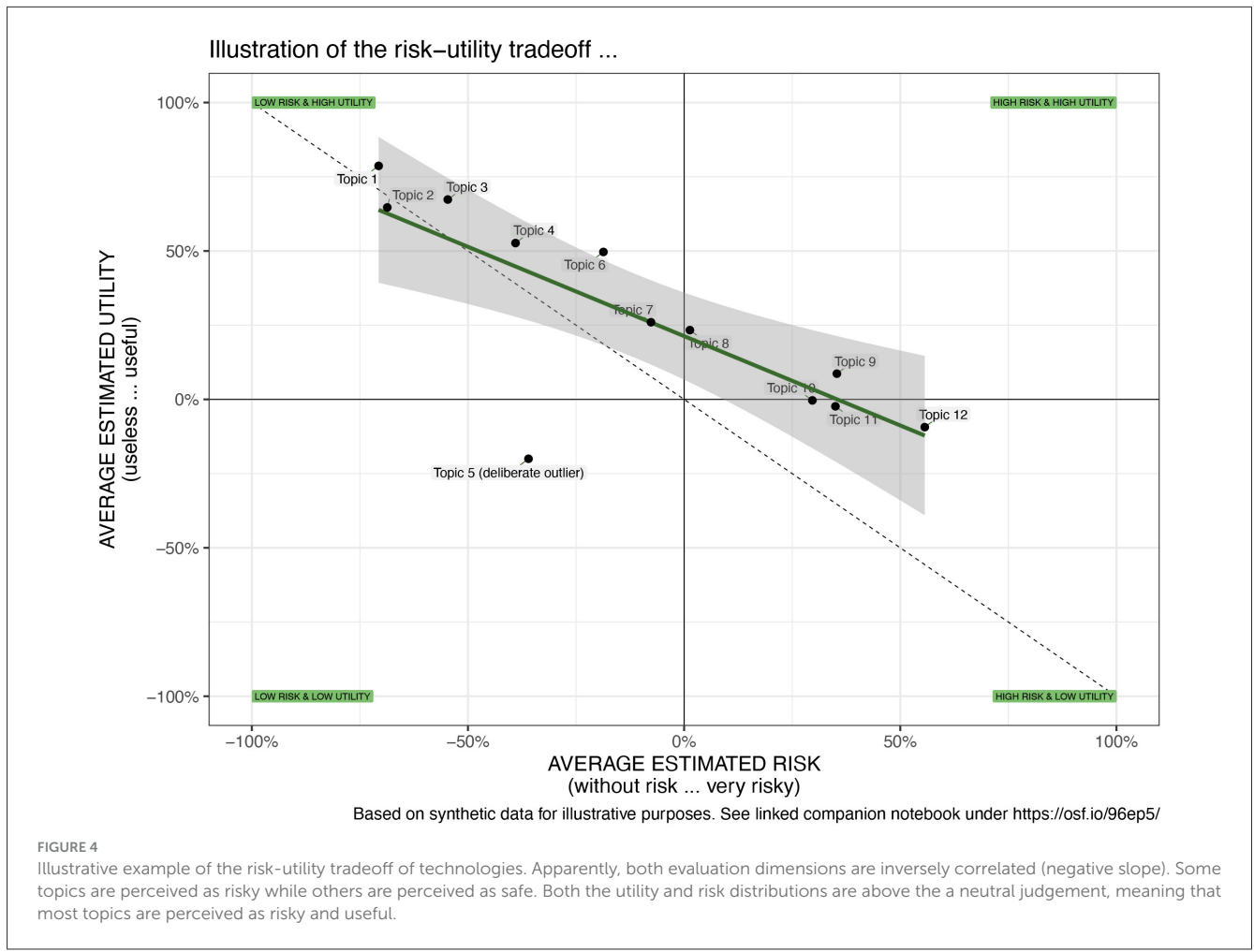

FIGURE 4
Illustrative example of the risk-utility tradeoff of technologies. Apparently, both evaluation dimensions are inversely correlated (negative slope). Some topics are perceived as risky while others are perceived as safe. Both the utility and risk distributions are above the a neutral judgement, meaning that most topics are perceived as risky and useful.

On the other hand, systematic biases can be avoided if the topics for the research space are compiled systematically. If possible, I recommend identifying an underlying factorial structure of the research space and exploring the research space systematically by querying 1 to $N$ topics for each linear combination of this space (i.e., latin square design). For example, if one wants to evaluate different forms of energy generation, one could first identify possible factors of an underlying design space of the topics (e.g., size, sustainability, risk, co-location with housing, ...) and their respective levels (e.g., ranging from small to large, not sustainable to circular, ...). Next, and based on the latin square method, suitable instances for each of the factor combinations can be identified and selected. This avoids that some areas of the underlying research space are over and others are under represented in the sample of topics. Hence, this approach reduces systematic bias in the data due to non-biased sampling of the topics.

Based on conducted studies, I suggest querying about 16–24 topics, to balance the expressiveness of the results with the length of the survey and to avoid the effects of both learning and fatigue. If more topics need to be queried, one can use *randomized sampling* of the queried topics: While many topics are in the survey, only a random subset is rated by each participant. Note however, that random sampling of technologies or topics may have unindented side-effects that may limit the validity of the study due to the risk of biased sampling.[1]

What suitable dimensions for the research space are, depends on the general research domain. As outlined above, a research space for energy conversation technology may build, for example, on the dimensions of degree of sustainability, price, size, or decentralization. A research space for medical technology (see Section 5) may build on the dimensions how invasive a technology is, how digital it is, whether it is used by patients or doctors. Beyond that one can also include other factors, such as when a topic or technology became public (cf. Protzko and Schooler, 2023).

Beyond the underlying factorial structure, the selected scenarios should otherwise be comparable. Participants should evaluate different instances of a technology and not hard to compare concepts. Of course, the scenario descriptions should be developed and iteratively refined to ensure comprehensibility for the participants and to facilitate the evocation of a mental model among the participants.

---

1 To mitigate this, one should build on a sufficiently large subset of technologies and a larger sample of participants. Further, one might consider suitable data imputation strategies.

## 4.2 Defining the topic evaluation variables

Next, the appropriate dependent variables for the assessment of the topics need to be identified. Of course, this depends on the selected research context and the targeted participants of the survey. For example, medical and biotechnologies often involve ethical considerations and personal values related to life, health, and body autonomy. Information and communication technologies (ICTs) influence and are influenced by values related to privacy, freedom of expression, and information accessibility. Hence, this article only provides some more general remarks on this selection: First, the article exemplifies the selection of variables by sketching three potential research questions. Secondly, it discusses how many and which items can be used for operationalization. Finally, it suggests how the reliability of the measurement can be checked.

For example, to study risk–benefit trade-offs and their relation to the willingness to accept or adopt a technology (Fischhoff, 2015; Brauner and Offermann, 2024), one might to query the *perceived risk*, the *perceived benefit*, and the overall acceptance or *willingness to adopt* a technology (Davis, 1989). This would allow to calculate a multiple linear regression (across the average topic evaluations) with the average risks and benefits of the technologies as independent variables and the willingness to adopt as dependent variable. For technologies that are not adopted by individuals (e.g., different types of power plants), an overall *valence* might be more suitable (Kim et al., 2007). In a different study, one might be interested in the *perceived sensitivity* and the *willingness to disclose* the information from various sensor types for personal life-logging (Lidynia et al., 2017) or workplace monitoring (Tolsdorf et al., 2022).

Suitable dependent variables can be adapted from other research models. For example, to evaluate a number of different mobile applications, one might refer to technology acceptance model (see above) and its key dimensions *perceived ease of use*, *perceived usefulness*, and *intention to use* or actual *use*. If the perception of risk and *benefits* (or utility) is of interest, one may consider *risk* and *benefits* as target variables: In one study, colleagues and I build on Fishhoff's psychometric model of risk perception (Fischhoff, 2015) to study risk-benefit tradeoffs in the context of Artificial Intelligence (AI) (Brauner et al., 2024): For a large number of developments and potential implications of AI, we wanted to explore if the overall evaluation is rather driven by the perceived risks or by the perceived benefits. Hence, we measured the overall evaluation as *valence* (positive—negative), the perceived *risk* (no risk—high risk), and the perceived *utility* (useless—useful).[2] Furthermore, one might study the *intention-behavior*-gap in different contexts.

In a recent study, an attempt was made to measure perceived expectancy, which refers to whether participants believed a presented development is likely to occur in the future (Brauner et al., 2023). However, no relationship to other variables in the study could be identified. This corroborates that forecasting seems to be difficult, especially for laypeople (Recchia et al., 2021).

---

2 Preliminary analysis of a still unpublished study on the perception of AI: https://osf.io/p93cy/.

The number of queried items for each topic should be limited. As the number of dependent variables for each topic increases the survey duration linearly, this can quickly lead to excessively long questionnaires. Hence, I am proposing to use single item scales for each relevant target dimension (Woods and Hampson, 2005; Rammstedt and Beierlein, 2014; Fuchs and Diamantopoulos, 2009). Consequently, I advise building on the existing research models and select the items that were identified as working particularly well in other studies (e.g., select the item with the highest item-total-correlation (ITC) from well-working scales).

In previous studies building on this approach, the number of target dimensions varied between two and seven. A number between three to five was working particularly well and should work for many contexts (e.g., to study the relationship between risk, benefit and acceptance, or intention and behavior).

Using a semantic differential for querying the dependent variables is suggested. These have metric properties and usually require low cognitive effort by the participants, as these items can usually be more easily interpreted, evaluated, and the appropriated response be selected (Messick, 1957; Woods and Hampson, 2005; Verhagen et al., 2018). Especially as the participants report on a larger number of scenarios and items, I suggest to keep the items and the response format as easy as possible.

## 4.3 Modeling the influence of user diversity

Finally, one should consider the choice of additional user variables that should be surveyed and related to the topic evaluations. Beyond the usual demographic variables, such as age and gender, this strongly depends on the specific research questions and context of the study. Hence, I can only provide some general ideas and remarks.

The first perspective of this approach facilitates the calculation of mean topic evaluations, for example, the mean valence or the mean risk attributed to the topics (see Figure 5). These calculated variables can then be considered as personality states (changing over time) or traits (relatively stable), and can be related to the additional user variables.

Hence, one should assume relationships between the newly calculated variables from the topic evaluation and the additional user variables. In the case of the study on the perception of AI, the average assessments across the topics (see Section 4.2) *valence*, *risk*, and *utility* were related with the participant's *age*, *gender*, *general risk disposition*, and *attitude toward technology*. If one aims at studying the intention-behavior gap regarding sustainable behavior (Linder et al., 2022), one may integrate, for example, constructs such as knowledge and attitude on climate change in the research model.

## 4.4 Balancing survey length and number of participants

Determining how many topics and how many target variables should be used is not trivial and depends on many factors. An obvious consideration is the number of included topics and dependent variables. Even if the repetitive query format facilitates efficient processing of the questionnaire (see above), both the number of topics and the number of dependent variables have an almost linear effect on the survey length.[3] Hence, the number of queried evaluation dimensions must be low. Otherwise the resulting questionnaire will be too long, resulting in reduced attention and increased dropout rates among participants. This consideration also depends on the sample and its motivation to participate: If participants are interested in the topic or are adequately compensated, more aspects can be integrated into the questionnaire. However, if participation is purely voluntary and the topic holds little interest for the participants, it is advisable to limit the number of topics and evaluation dimensions.

Defining the required sample size depends on the desired margin of error for the measurements and the empirical variance of the dependent variables used in the technology assessments. The required sample size $n$ can be calculated using the formula (Häder, 2022; Field, 2009): $n = (\frac{Z \cdot \sigma}{E})^2$ where $\sigma$ is the (unknown) standard deviation of the population, $Z$ is the critical value for the desired confidence level (for example 1.65 for a 90% confidence interval or 1.96 for a 95% confidence interval, with the latter being commonly used in the social sciences), and $E$ is the targeted margin of error in units of the dependent variable scale (e.g., 0.5 if a deviation of ±0.5 unit from the true mean is acceptable on a scale ranging from −3 to +3). The variance $\sigma^2$ can be estimated from prior research, suitable assumptions, or a pilot study. Both the desired confidence level ($Z$) and the acceptable margin of error ($E$) depend on the research goals and required precision and need to be defined by the researcher. Exploratory studies might accept higher margins of error, while confirmatory studies typically demand lower error ranges. It is important to note that if only a subset of topics is randomly sampled, this would increase the required sample size.

Based on experience, I recommend gathering at least 100 participants per topic evaluation. This sample size has yielded a margin of error of about 0.25, given the measured variance and a 95% confidence interval. By considering these factors and calculating the sample size accordingly, researchers can ensure that their findings are both statistically valid and meaningful within their research context.

## 4.5 Visualizing the outcomes

A particular advantage of this approach is that the results of the technology assessment can be clearly and accessibly presented in addition to the various possible statistical analyses.

I especially suggest the use of 2d scatter plots, which can illustrate the relationship between two dependent variables across themes (such as risk on the $x$-axis and utility on the $y$-axis), or of one dependent variable across two user groups (such as the risk assessment between laypeople on the $x$-axis and experts on the $y$-axis).

Since many possible visualizations can be made based on the number of different dependent variables or different groups of participants, one should focus on the most relevant ones. Here, of course, it is advisable to first select dimensions that are particularly relevant from the research question or a theoretical perspective (such as the aforementioned risk–benefit trade-off; even if the variable valence is used for calculations but not illustrated). Alternatively and especially for more exploratory studies, one can also display pairs of variables that have a particularly strong or weak relationship with each other. Note that readers will profit from good illustrations and clear annotations what the figure conveys. Hence, the axis, quadrants, and regression lines should be labeled clearly and readers should be guided through the interpretation of the diagram.

---

[3] For example, if the number of dependent variables is increased from three to four the expected survey duration grows by 33%.

## 4.6 Drawbacks, challenges, and outlook

Besides advantages and insights, each method in the social sciences has its disadvantages and limitations. The following section discusses the limitations and challenges of the micro-scenario approach. Suitable alternatives are suggested afterwards.

Two (deliberate) limitations of the micro-scenarios are the brevity of the scenario narrative and the concise assessment using only a few response items. The consequence of this terseness is potentially less precise evaluations, likely contributing to greater variance in the data.

Since the scenarios cannot be presented in greater detail (compared to single scenario evaluations), the mental models of the participants—and these mental models are ultimately evaluated—can differ substantially and may be oversimplified. Of course, this is not necessarily a disadvantage if the research goal is the quantification of the affective evaluation of various topics (Finucane et al., 2000; Slovic et al., 2002). Nonetheless, possible alternatives should be considered and measures should be taken to mitigate this drawback of this approach.

If the topic evaluations are queried on single items scales, one cannot calculate reliability measures for the constructs [e.g., Cronbach's alpha ($\alpha$) or McDonald's Omega ($\omega$) as common measures for internal reliability]. Additionally, given the vast number of dependent variables collected ($n \times m$, represented by the product of the number of topics $n$ and the number of outcome variables $m$), a detailed analysis of each variable's distribution and associated characteristics (for instance, normality and unimodality) for each topic is impractical. The use of single-item scales by itself is doable, if one has the reasonable assumption that the measured construct is unidimensional, well-defined, and narrow in scope (Rammstedt and Beierlein, 2014; Fuchs and Diamantopoulos, 2009; Ang and Eisend, 2017; Woods and Hampson, 2005). In this respect, one should have sensible prior assumptions regarding the planned dependent variables or carry out accompanying studies to test these.

While the internal consistency cannot be calculated, one can calculate other reliability measures, such as the intraclass correlation coefficient (ICC). This measures if the raters (i.e., the participants of a study) agree with their ratings on each single-item scale across the different queried topics (Cicchetti, 1994). Consequently, higher ICCs would indicate a consistency in the evaluations, with some technologies or topics rated as higher and others as lower. But although high consistency is important for the construction of a psychometric scale, it cannot be assumed for technology assessment: For example, society has no unanimous opinion on technologies such as nuclear power (Slovic, 1996) or wind power (Wolsink, 2007). In this respect, different opinions influence the measured ICC.

A limitation is that the interactions between a topic or a set of topics and the participants cannot easily be identified or interpreted. If the results suggest specific outliers or interactions, one is advised to re-evaluate the specific technologies using alternative methods for mental model extraction (such as topic-specific surveys or interviews) that allow more robust measurements in exchange for less queried topics.

When evaluating scenarios, it is essential that a good scenario description evokes a clear mental model in the participants and that they can evaluate it as accurately as possible with regard to the research question. Even more than in studies with one or a few scenarios, in the micro-scenario approach researchers must ensure that the scenarios are formulated concisely and that the response items can be clearly interpreted by the participants. Due to the breadth of topics covered in a micro-scenario study, intensive pretesting of the scenario descriptions, the evaluation metrics and the tools used is essential.

One solution to mitigate these issues could involve providing lengthier and more detailed scenario description alongside more comprehensive response items. However, maintaining the questionnaire's duration constant would necessitate a transition to a between-subjects design or the partitioning of scenarios and their evaluations across multiple studies. In an extreme scenario, a cumulative evaluation could be constructed through a meta-analysis across numerous studies with a similar structure. Such measures would undeniably enrich the validity of the results but at the cost of requiring substantially more participants and resources. Hence, this would annihilate the advantages that the micro-scenario methodology offers, such as a within-subject measurement, efficiency, and rapid data collection.

As noted earlier, studies suggest that the relationships between survey items and constructs can be distorted by lexical biases, such as word co-occurrences (Gefen and Larsen, 2017). While micro-scenarios alone won't fully resolve this issue, they can help explain and mitigate its effects. Unlike abstract or generalized survey items, micro-scenarios present specific, contextualized situations. This specificity may reduce the impact of lexical similarity, which can otherwise skew responses due to the proximity of wording rather than reflecting genuine differences in perception, particularly when comparing studies from different contexts but with the same outcome variables. By integrating multiple scenarios into a single comprehensive survey, micro-scenarios enable the evaluation of a wide range of technologies and concepts simultaneously. This approach captures more nuanced insights and reflects a broader spectrum of user experiences, reducing the reliance on potentially biased single-topic constructs. Furthermore, micro-scenarios facilitate reflexive measurement across different technologies or topics, better accounting for individual differences in technology perception. This goes beyond surface-level responses, revealing deeper patterns in how people relate to technology, thus addressing limitations in traditional methods. In summary, micro-scenarios may reduce lexical biases and enhance the robustness of technology acceptance assessments by complementing traditional methods with a more contextualized, comprehensive, and nuanced approach to understanding public perception.

## 5 Application example

To make the application and potential outcomes of this method more tangible, I will present the structure and results of an study on the acceptance of medical technology I contributed to. Detailed information on the goal of the article, its methodological approach, sample, and results can be found in the corresponding article (Brauner and Offermann, 2024). The aim of the study was to investigate how various medical technologies are assessed in terms of perceived risk and benefits, as well as a general valence evaluation. Additionally, the study sought to determine which of the two predictors—perceived risk or benefits—has a stronger influence on valence, and whether user factors affect this evaluation.

Initially, we compiled a list of 20 different medical technologies in workshops, ensuring a balance between older and newer, as well as invasive and non-invasive technologies. The technologies ranged from adhesive bandages and X-rays to mRNA vaccines. We then had these technologies evaluated by 193 participants using the assessment dimensions of perceived risk, perceived benefit, and general valence (ranging from negative to positive).

The results are 3-fold:

First, in general and across all queried technologies and participants, medical technologies are perceived as rather safe (Risk = −44.5%) and useful (Utility = 48.4%) by the participants. Similarly, the overall attributed valence—that is how positive or negative the participants evaluate the technology—is rather positive (Valence = 49.0%). Figure 5 illustrates the distribution of the evaluations.

Second, when the overall assessments of the topics were interpreted as an individual difference (Perspective 1, see Section 3), the results suggest that the valence toward medical technology is linked by individual differences, with caregiving experience and trust in physicians emerging as significant predictors.

Third, Figure 6 illustrates the risk-utility tradeoffs and the negative relationship between perceived risk and perceived benefits ($r = -0.647$, $p = 0.02$). It shows that technologies like "home emergency call button" and "plaster cast" are both highly useful and carry low perceived risk, whereas "robotic surgery" and "insulin pumps" are seen as useful but carry higher perceived risks. Finally, the novel "mRNA vaccines" are perceived as relatively high risk and low utility compared to other technologies in this study, which might reflect public skepticism or misinformation during the survey period. Furthermore, a regression analysis suggest that much of the variance in valence ($R^2 = 0.959$) is predicted by utility ($\beta = 0.886$) and to a lesser extend by the perceived risk ($\beta = -0.133$). Overall, this chart provides a visual representation of the public opinion on various medical technologies and how these are perceived in terms of their risks and benefits. It helps to identify which technologies are most favorably viewed (top-left quadrant) and which are viewed with skepticism (bottom-right quadrant). It can inform policymakers, healthcare providers, and technology developers on areas where perceptions of risk and utility may need to be addressed, which could be crucial for adoption strategies, communication plans, and further research.

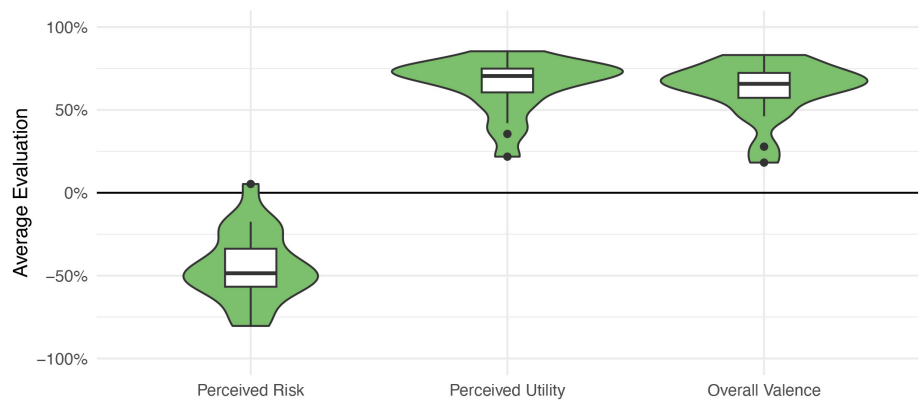

FIGURE 5
Average evaluations of 20 medical technologies by 193 participants showing that most medical technologies are seen as low risk, beneficial, and positive. Adapted from Brauner and Offermann (2024).

# 6 Conclusion

Overall, the presented approach enables a superordinate comparison and visualization of the acceptance and perception of a broad variety of technologies and concepts (context-specifically or cross-contextually) on different measures.

The interaction between technology and people and their values is complex and multifaceted. Some technologies can be directly compared based on public perception and mental models, particularly those within the same domain or serving similar functions. Others may require more nuanced, context-specific evaluations. This section discusses key insights, advantages, and limitations of this approach.

In general, the *approach is pragmatic* and provides an accessible comparative overview of the acceptance and perception of technologies or technology-related concepts by integrating the evaluation of *many topics* (i.e., diverse technologies in a specific or various contexts) in a single comprehensive study. This entails many advantages as it *can inform various target groups* about potentially critical issues. For example, for technology developers and researchers, this approach provides ideas and starting points to improve and develop critical technologies alongside future users' needs and perceptions. For social scientists, insights from this approach enables them to derive recommendations regarding information and (risk) communication to address future users' needs and requirements. Finally, the insights of this approach can also be used by policymakers as the basis for decision-making for governance, as it provides information about what has to be controlled better, where priorities should be set within the development and realization of innovation technologies and applications, and where citizens need more information and involvement.

Beyond the comparative overview the approach offers methodological benefits: First of all, the approach enables the transformation of the topic evaluations into *visual cognitive maps*. Herein, the different topics from the same domain can be viewed in their spatial relation to the other topics and their absolute placement. Furthermore, the relationships between the queried target variables can be statistically analyzed, for example, by interpreting their correlations, slopes, and intercepts. Various perspectives can be studied (partially based on the visualizations) within the introduced approach: A *contextual analysis* provides insights on how different topics are related to each other, and reveals potential outliers. Furthermore, the placement of the dots (as the *mean evaluations* of each topics) on the axes show how the topics are placed and perceived (e.g., rather risky or not). The *dispersion*, that is, the distribution of the dots across the scales, indicates the consistency of the evaluations and shows whether they represent uniform or rather diverse attributions. Further, *correlations* between the attributions can be analyzed and show how strong different evaluations are related across the topics. Additionally, different *intercepts on the axes* and thus the position of the topics can also be analyzed and interpreted. If three or more variables are evaluated per topic and one is a dedicated target variable, the *degree of explained variance* can be interpreted by means of regression analyses, to inform how uniform the topic evaluation is across all topics. Regression analyses also inform which factors have the *strongest influence* on the target variables (such as valence). These results can then be used to, for example, derive adequate and tailored communication strategies. Finally, as with other approaches, the overall evaluations per participants can be *linked to other responses from the participant*, such as their demographics, attitudes, beliefs, or reported behaviors. In this regard, the introduced mapping and visualizations of the evaluations can also be realized to compare different sub-samples depending on specific variables (e.g., age groups, low vs. high technology expertise).

In addition, the article provides *practical tools* in terms of specific recommendations and R code alongside the methodological concept, which will help easily use and directly apply the presented approach in future research.

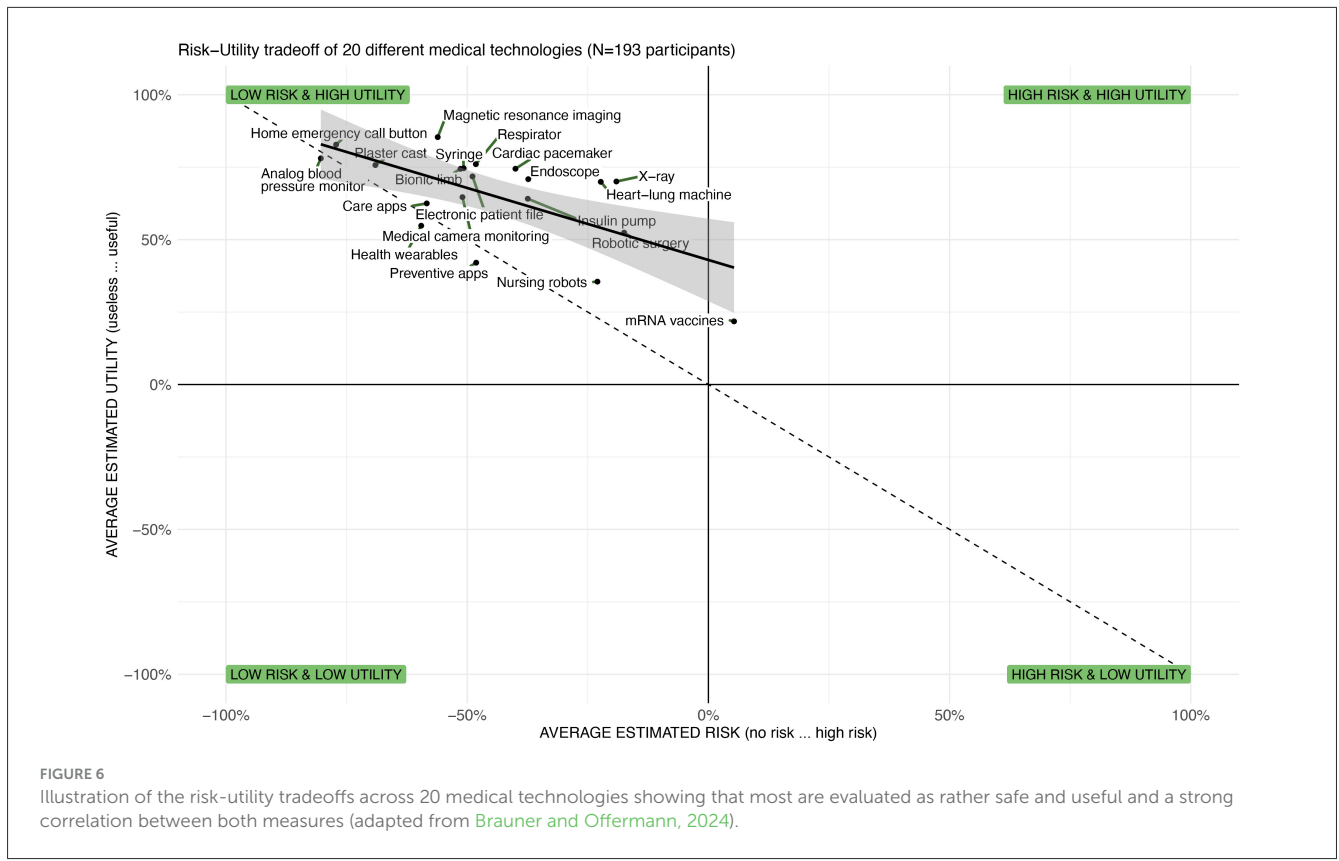

FIGURE 6
Illustration of the risk-utility tradeoffs across 20 medical technologies showing that most are evaluated as rather safe and useful and a strong correlation between both measures (adapted from Brauner and Offermann, 2024).

Summarizing the methodological advancements of the micro-scenario approach, the dual complementary perspectives offer three significant benefits. First, they facilitate the modeling of individual differences through reflexive measurement across various technologies or topics. Second, they provide valuable insights for developers, researchers, and policymakers by analyzing the spatial positioning of the topics to identifying critical issues in technology perception. Third, this enables the identification of acceptance-relevant factors crucial for tailoring technology to better meet human needs.

## Data availability statement

The original contributions presented in the study are included in the article/supplementary material, further inquiries can be directed to the corresponding author. An executable R notebook that extends the example code provided in this article is publicly available https://github.com/braunerphilipp/MappingAcceptance. All data used for generating the example is available at the repository.

## Author contributions

PB: Conceptualization, Data curation, Formal analysis, Funding acquisition, Investigation, Methodology, Project administration, Resources, Software, Supervision, Validation, Visualization, Writing – original draft, Writing – review & editing.

## Funding

The author(s) declare financial support was received for the research, authorship, and/or publication of this article. This work was funded by the Deutsche Forschungsgemeinschaft (DFG, German Research Foundation) under Germany's Excellence Strategy—EXC-2023 Internet of Production—390621612. Open access funding provided by the Open Access Publishing Fund of RWTH Aachen University.

## Acknowledgments

This approach evolved over time and through several research projects. I would like to thank all those who have directly or indirectly, consciously or unconsciously, inspired me to take a closer look at this approach and who have given me the opportunity to apply this approach in various contexts. In particular, I would like to thank: Julia Offermann, for invaluable discussions about this approach and so much encouragement and constructive comments during the final meters of the manuscript. Martina Ziefle for igniting scientific curiosity and motivating me to embark on a journey of boundless creativity and exploration. Ralf Philipsen, without whom the very first study with that approach would never have happened, as we developed the crazy idea to explore the benefits of barriers of using "side-by-side" questions in Limesurvey. Julian Hildebrandt for in-depth discussions on the approach and for validating the accompanying code. Tim Schmeckel for feedback on the draft of this article. Felix Glawe, Luca Liehner, and Luisa Vervier for working on a study that took this concept to another level. Of course, I would also like to thank my two referees and the editors. They were very accurate in their identification of weaknesses and ambiguities, and their very constructive feedback really contributed to the strengthening of the article. I express my gratitude to my son Oskar, who—while teething and sleeping on my belly at night—allowed me to write substantial portions of this article. I utilized Large Language Models (LLMs), specifically OpenAI's GPT-3.5 and GPT-4o, for assistance in editing and R coding. For writing assistance, typical prompts included requests such as, "I am a scientist writing an academic article. Can you edit the following paragraph? Please explain the changes you made and the rationale behind them." For coding support, prompts included, "I am coding in R and have two data frames, A and B. How do I merge these using the unique ID in `tidyverse` syntax?" Importantly, LLMs were not used to generate (or "hallucinate") any original content for the manuscript.

"Your methods turn out to be ideal. Go ahead."—Quote from one of my last fortune cookies. No scientific method of the social sciences alone will fully answer all of our questions. I hope that this method provides a fresh perspective on exciting and relevant questions.

If you have carried out a study based on this method, please let me know with the research context and the investigated variables and I will document it on the project's page (https://github.com/braunerphilipp/MappingAcceptance).

## Conflict of interest

The author declares that the research was conducted in the absence of any commercial or financial relationships that could be construed as a potential conflict of interest.

## Publisher's note

# Appendix: practical tools: implementing and analyzing micro scenarios

This appendix provides practical tips for implementing micro scenarios in surveys and analyzing the resulting data. An executable R notebook[4] offers a comprehensive example, including code for data transformation, analysis, and visualization.

## Implementing micro scenarios in survey tools

Many survey tools simplify the creation, data collection, and analysis of online questionnaires, reducing the need for manual input.

For example, the *side-by-side* question format (available in tools like Limesurvey and Qualtrics) displays topics and their response items as rows in a table. While easy to process, this format requires all items to be displayed on the same page, which may overwhelm participants or be difficult to view on small screens.

Some tools like Qualtrics offer advanced options such as *Loop & Merge*, which generates repeating blocks based on a data table (e.g., topic titles and descriptions). The tool iterates through all or a subset of topics, presenting them with consistent formatting. Survey data is stored in a structured format, with response variables named systematically (e.g., `aN_matrix_M`, where `N` is the topic number and `M` the dimension).

## Data analysis

Standardized variable names, like those generated by *Loop & Merge*, allow for systematic and automated data transformation. Below, I provide R code examples using the `tidyverse` package (https://www.tidyverse.org/). Other software can also be used.

### Rearranging survey data from wide to long format

Survey data must be reshaped from wide format (one row per participant, as in Figure 2) to long format (one value per row for each participant, topic, and evaluation dimension). Listing 1 demonstrates this transformation using `pivot_longer`. Additionally, survey responses (e.g., 1–7 scales) are rescaled to a percentage format ranging from −100% to +100%. Other variables, such as demographics, are neglected but will be added at a later stage.

Listing 1  Convert survey data to long format (one row per observation.

```
long <- surveydata %>%
dplyr::select(id, matches("a\\d+
\\_matrix\\_\\d+")) %>%
tidyr::pivot_longer(
cols = matches("a\\d+\\_matrix\\_\\d+"),
```

---

[4] https://github.com/braunerphilipp/MappingAcceptance

```
names_to = c("question", "dimension"),
names_pattern = "(.*)_matrix_(.*)",
# Separate topic and evaluation
values_to = "value",
values_drop_na = FALSE) %>%
dplyr::mutate( dimension = as.numeric
(dimension) ) %>% # readable dims
dplyr::mutate( dimension = DIMENSIONS
[dimension]) %>%
dplyr::mutate( value = -(((value - 1)/3)
- 1)) # rescale [1..7] to [-100%..100%]
```

### Calculating grand means for dimensions

In Listing 2, the grand mean for each assessment dimension is calculated across all topics and participants.

Listing 2  Calculate grand mean for each assessment dimension.

```
byDimension <- long %>%
dplyr::group_by( dimension ) %>%
dplyr::summarise( mean = mean(value,
na.rm = TRUE),
sd = sd(value, na.rm = TRUE), .groups="drop")
```

### Research perspective 1: calculate average topic evaluations as individual differences

Listing 3 shows how to pivot the data back to wide format and calculate the average topic evaluations for each participant. After pivoting, participants' topic evaluations are aggregated (e.g., by mean or median). The resulting data has one row per participant and columns for the average evaluation. These results can be merged with original survey responses using `left_join` to explore relationships with other variables (see Section 4.3).

Listing 3  Calculate average topic evaluations for each participant.

```
byParticipant <- long %>%
tidyr::pivot_wider(
names_from = dimension,
values_from = value) %>%
dplyr::group_by(id) %>%
dplyr::summarize(
across(
all_of( DIMENSIONS ), # Select evaluation
dimensions
list( mean = ~mean(., na.rm = TRUE),
median = ~median(., na.rm = TRUE),
sd = ~sd(., na.rm = TRUE)),
.names = "{.col}_{.fn}" # Define schema for
column names
), .groups="drop"
) %>%
dplyr::left_join(surveydata, by="id")
```

## Research perspective 2: calculate average topic evaluations

Listing 4 shows how to calculate the average assessments for each topic by summarizing data using the arithmetic mean and standard deviation. As shown in Figure 3, the data now contains one row per topic with two columns (mean and SD) for each dimension. This topic-related data can now be analysed or visualised.

Listing 4  Calculate average evaluations for each topic.

```
byTopic <- long %>%
tidyr::pivot_wider(
names_from = dimension,
values_from = value) %>%
dplyr::group_by(question) %>%
dplyr::summarize(
across(
all_of(DIMENSIONS), # Select evaluation dimensions
list(mean = ~mean(., na.rm = TRUE),
sd = ~sd(., na.rm = TRUE)),
.names = "{.col}_{.fn}" # Define schema for column names
), .groups="drop")
```